# Predicting Dynamic Replication based on Fuzzy System in Data Grid

Mahnaz Khojand[1], Mehdi Fatan Serj[2], Sevin Ashrafi[3], Vahideh namaki[4]

[1] *Islamic Azad University of Zanjan, Zanjan, Iran*
[2] *Departement of Computer Computer science and security of mathematic of Rovira i Virgili University, Tarragona, Spain*
[3] *Departement of Computer Engineering, Islamic Azad University of Arak, Arak, Iran*
[4] *Departement of Computer Engineering, Islamic Azad University of shabestar, shabestar, Iran*

*Abstract*

Data grid replication is an effective method to achieve efficient and fault tolerant data access while reducing access latency and bandwidth consumption in grids. Since we have storage limitation, a replica should be created in the best site. Through evaluation of previously suggested algorithms, we understand that by blind creation of replications on different sites after each demand, we may be able to improve algorithm regarding response time. In practice, however, most of the created replications will never be used and existing resources in Grid will be wasted through the creation of unused replications. In this paper, we propose a new dynamic replication algorithm called Predictive Fuzzy Replication (PFR). PFR not only redefines the Balanced Ant Colony Optimization (BACO) algorithm, which is used for job scheduling in grids, but also uses it for replication in appropriate sites in the data grid. The new algorithm considers the history usage of files, files size, the level of the sites and free available space for replication and tries to predict future needs and pre replicates them in the resources that are more suitable or decides which replica should be deleted if there is not enough space for replicating. This algorithm considers the related files of the replicated file and replicates them considering their own history. PFR acts more efficiently than Cascading method, which is one of the algorithms in optimized use of existing replicas.

*Keywords*: Data Grid, Cascading method, Fuzzy Systems, Cluster

# 1   Introduction

Grid system is one of the various kinds of distributed systems that have been designed until now. Grids can be classified into computational Grids and Data Grids [1]. Computational grids are developed for managing and handling computational tasks and Data Grids manage huge data files and data sharing [2]. We focus here on the data distribution aspect of a grid. With today's great projects and large datasets, Data Grids need large computational power and data storage resources [3]. Data Grids use the storage resources that are distributed in the whole world for the evaluation of these projects and the management of large amount of data that these projects need [4, 5, 6]. Because of geographical distribution of data resources, data transmission is unavailable. Data transmissions, especially long and large ones, cause time latency, bandwidth consumption and consequently decrease in performance. Data replication is one of the solutions for this problem [2, 7, 8, 9, 10, 11]. Concept of replication means making several copies of one identical data on different sites and in different places. Through such an action, availability along with reliability of the system increases. Furthermore, it helps to create load balancing in the system [2, 18].

Replication is done in two different ways; static and dynamic. In static mode, copying is independent from client's behaviour and is unchangeable. However, in dynamic copying mode, changes that are resulted from the environment have direct influence on production places and on their creation or deletion [13]. Since grid structure is always changing, we make use of dynamic methods to copy. If we copy all data in all sites, we will have the highest access-rate. However, because of the storage limitation, this kind of replicating is not possible. The most important issue in replicating is introducing proper strategies to determine the best time and the best place of creation and deletion of the replicas.

Since we face memory limitation, one of the most important challenge regarding Grid environment optimization is to increase the percentage of average accessibility to replicas and consequently to optimize the memory which is used by replicas in a way that when a replica is created and takes up a place, we should make the most use of it. Through evaluation of previously suggested algorithms, we understand that by blind creation of replicas on different sites after each demand, we may be able to improve algorithm regarding response time, but in practice, most of the created replicas will never be used and existing resources in Grid will be wasted through the creation of unused replicas.

Our proposed algorithm Predictive Fuzzy Replication (PFR) is a new dynamic method in Multi-Tier Data Grid Environment. In this paper, we redefine the Balanced Ant Colony Optimization (BACO) algorithm [12], which is used for job scheduling in Computing grids, and use it for replication in appropriate sites in the data grid. In BACO, each job is an ant and ants are sent to search for resources. A job in BACO is a file; a resource is a node in our method. PFR considers the history usage of files and tries to predict future need and pre replicates them in the resources that are more suitable or decides which replica should be deleted if there is not enough space for replicating. This algorithm considers the related files of the replicated file and replicates them considering their own history. Creating copies in appropriate sites, before users' requests, causes access delay decreases significantly. Contrary to most of the previous algorithms, which only consider leaf nodes, PFR is used all over the Grid. Since it is impossible to examine all nodes in PFR due to the wide amount of nodes, in this article, we will examine algorithm rather than discussing the nodes on cluster levels.

Our proposed algorithm, PFR, tries to create more optimized replications. PFR acts even more efficiently than Cascading method, which is one of the algorithms in optimized use of existing replicas.

The rest of the paper is organized as follows. In section 2, we report some related research work on replication strategy in Data grid. Section 3 introduces our algorithm, results are presented in section 4, and section 5 offers future works and concludes the paper.

# 2   Related Work

Ranganathan in [13] introduced five different replication strategies for three different kinds of access patterns.
These strategies are:
- No Replication or Caching: No replication is created in any node.
- Best client: The client that has the maximum request for a file is identified (Best client) and that file is replicated on it.
- Cascading: When the request number for a file exceeds the threshold, the file is replicated at the next level.
- Caching and cascading: This strategy acts like cascading method with this difference that the requested file is replicated in the client node too.

- Fast spread: after file demand, the file is copied in all the existing nodes on the path from the root to the client.

The access patterns that they introduced are:
- Temporal Locality: Recently accessed files with high probability will to be accessed again.
- Geographical locality: Files recently accessed by a client with high probability will be accessed by nearby clients.
- Spatial Locality: Files related to the recently accessed file will be accessed too.

Our proposed method uses spread strategy and considers temporal, geographical and spatial localities.

BHR [26] determines network layers in form of various regions. Bandwidth among sites inside the same region is much higher than bandwidth among different regions. Therefore, if the requested file is fetched from a site which is in another region, the response time will highly increase. If when a file is requested, the file doesn't exist in the respective site, in case the site has enough space, the file will be copied in it. Otherwise, if the file exists in another site in that region, the file won't be saved. If the file isn't found in that region in that region, those files which have been copied more than once in the sites in that region are identified and then, the files that have recently been used less are deleted until we have enough space to copy the new file. MBHR [3] is a modified version of BHR which considers geographical and temporal locality as well and tries to create all copies in one region and in the sites that are being accessed more. We have used spatial locality in addition to temporal and geographical locality and clustering concept instead of regional network.

Sonali Warhade in [25] uses the graph topology to form the grid system. In his method, through raising the modified BHR (MBHR) methodology, he has a tendency to project a dynamic algorithmic program for data replication in data grid system. This algorithmic program uses variety of parameter for locating replicating appropriate web site wherever the file is also needed in future with high likelihood. The algorithmic program predicts future wants of replicated appropriate grid web site square measure supported file access history.

Peter proposed a method in [10], the file is divided in to blocks. If there is a need for replication only the needed block is replicated in the site so it saves the storage space and makes the parallel access possible to replica. This method does not consider spatial locality. The authors in [20] studied data replicating in multi-tier data grid and suggested two simple bottom-up and aggregative bottom-up replicating algorithm. The main idea of these methods is that they identify the popular files and replicates those files in closest place to the client, like in its parent. In this way, each node is considered like a server for its children. The copy trend in this method is a bottom-up process. A threshold is considered in SBU. If the demand rate for a file by a client exceeds that limit, a copy will be created in the nearest place on a higher level than the client node. SBU encounters problems if it takes all the system in to consideration. ABU is offered as a solution to this problem. Since it is logical, when a file is demanded repeatedly by most of the children of a node, a copy is created in the parent. ABU aggregates the history of all accesses to files from all nodes on each level, finds their relationships with each other and on that basis, does the copying until it gets to the root. According to the response time, ABU is better than SBU and fast spread. In SBU and ABU, files access number is a scale for identifying popular files. One of the related problems is that one file was perhaps visited frequently before, but although there is no request for the file now, the copy of the file is produced on the basis of its history. This problem was discussed in [21] and the solution is an algorithm called Last Access Largest Weight. The most important point in LALW is that it gives records different weights, on the basis of their age. In fact, LALW gives higher weights to the new records.

Tao Wang in [26] proposed a method, named FLSDR, a theoretical model of access latency optimization with replication, which complements the blank space. Subsequently, he proposed a carefully designed dynamic replication consisting of three algorithms. These logarithms are as follows: replica selection algorithm, replica layout algorithm and replica replacement algorithm. Replica selection algorithm selects the optimal replica with a hierarchical time cost based on the derivation of the theoretical model. Replica layout algorithm selects the optimal node for placing the replica based on the spatio-temporal locality of data access. Replica replacement algorithm, in which the fuzzy logic system is introduced originally, deletes replica when the available storage space is insufficient. FLSDR achieves better performance in comparison with other algorithms in terms of mean job execution time, computing resource usage, number of data scheduling between clusters and number of replicas. PHFS [2] is a dynamic replication method which predicts replication places before demand by using spatial locality and spread strategy. PHFS uses data mining methods to find the relationship between files along with the percentage of relationship among them. When a file is demanded for the first time, in addition to the file itself, related files, are replicated in lower levels. This is done in a way that the files, which have lower dependency, are only replicated in higher levels but files with higher dependency are also replicated in lower levels. In case of later demands, replication is done base on the priority attributed to files regarding their usage rate in the past. In our own algorithm, we have also tried to make accessibility more optimized through replicating the related files. The difference, however, is that we consider all over the grid in our algorithm and always apply algorithm on all nodes and files. To replicate related files, in addition to the percentage of their relationship to the main selected file, we also consider the characteristics of files in order to be replicated in lower levels [2].

Through using ACO algorithm, BACO algorithm has offered a method to schedule jobs. Supposing that each job is an ant, the ants are sent to search for resources. While trying to reduce computation time to run jobs through local and global updates, this algorithm controls load balancing in Grid resources and uses a matrix to show computation power of each resource for each job [12]. We generalize this algorithm from Computing Grid to Data Grid, and we use this matrix to determine the best place to replicate files.

Grid can have different architectures. One of the popular architectures is multi-tier structure. This architecture offers a simple structure to apply replicating methods. Multi-tier architecture in a data grid is a tree-like structure in which the nodes connect hierarchically and according to tree topology limits.

Multi-Tier architecture was first proposed by MONARC project [14] for the Large Hadron Collider (LHC) [15]. European Organization for Nuclear Research (CERN) [16] is working on LHC. Multi-Tier architecture is 5-layer architecture on whose 0 layer, CERN is situated, all data are placed in CERN. Its first layer includes national centers. The second layer is sub-group of the first and includes regional centers. The third layer consists of industrial centers and work groups. Desktops and end users are in the fourth layer. We have used this architecture for implementation of our proposed algorithm.

## 3    Predictive Fuzzy Replication

In this section, a new dynamic replication method in the multi-tier data grid called Predictive Fuzzy Replication (PFR) will be described. PFR considers the history usage of files, files size, the level of the sites in multi-tier architecture and free available space for replication and tries to predict future needs and pre replicates them in the resources that are more suitable. In this paper we have used CERN [16] Multi-Tier data grid for the implementation of PFR method.

Due to great number of nodes in grid, instead of discussing sites individually, we discuss clusters and instead of considering all available sites, we will only consider cluster headers related to each cluster. Cluster header is a node, which has a higher computational power and storage capacity, compared to other nodes (Figure 1). Each cluster has only one cluster header, which keeps the information related to the sites that exist in its own area along with places for present replications. Each cluster header has a replication manager. The replicating manager is responsible to control replications inside the cluster and in case replication is needed, it chooses the best site in its respective cluster to accept the replication.

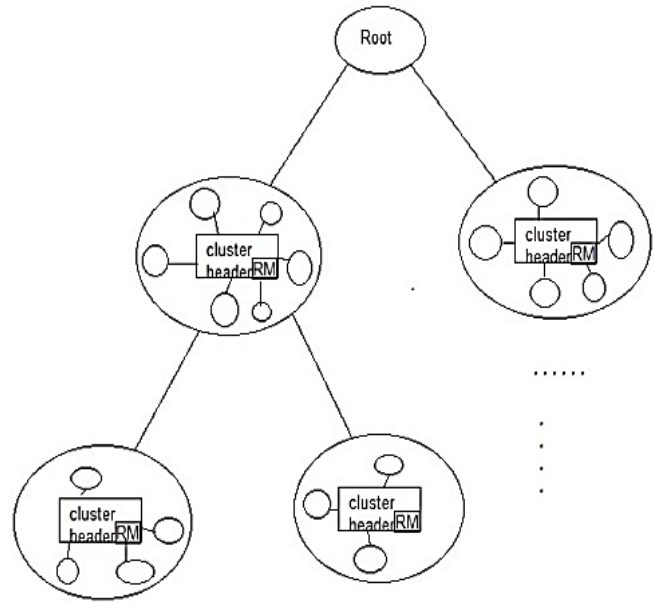

**Figure 1:** Clustering in data grid

In order to cluster the nodes, clustering algorithms [25] or graph partitioning algorithms [24] can be used. We have used a simple algorithm here in a way that nodes whose distance from each other is closer than the limit $\alpha$ and which do not belong to other clusters, are grouped in one cluster. $\alpha$ is a user defined value here.

In other words, if we show the existing sites with an index of S, M is defined as Universal Set, which contains all other sites. R is a set of nodes that belong to cluster A ( *Eq.* 1, 2). Consider Table (1) for more information.

**Table 1:** The list of Eq. 1,2. Parameters.

| | |
|---|---|
| M | Universal Set |
| S | An Existing sites are shown with an index of S |
| R | set of nodes that belong to cluster A |
| X | a node in the grid |
| A | is a node in cluster A |
| $cl_A$ | denotes cluster A |

$$M = \{S_1, S_2, \ldots, S_n\} \quad (1)$$

$$R = \{x \mid x \in cl_A \land (|a - x| < \alpha \land x \notin (M - cl_A)\} \quad (2)$$

In Multi-Tier data grid nodes that exist in upper levels of hierarchical tree, have higher data storage capacity and computing power than lower nodes. We have considered all three access patterns with temporal, geographical and spatial locality in our method. We take all our data as read only, that is, we do not need to update the replications.

To find the best place for replication in data grid, PFR uses a method similar to the method in BACO, the way used to find the best resource for running the job. In addition, PFR uses Replica Indicator (RI) matrix for showing the relation between the nodes and the files, like BACO that uses a matrix for showing the relation between the resources and the jobs.

The rest of the section is organized as follows. In 3.1, we describe how to make replica indicator matrix. 3.2 describes making Dependency matrix. , PFR algorithm are presented in 3.3, and 3.4 offers an architecture for PFR and 3.5 shows an example of our proposed algorithm.

### 3.1 Making Replica Indicator matrix

In the suggested algorithm, to make matrix Replica Indicator (RI), if we have m nodes and n data (here, node means cluster header), we will have a $m \times n$ matrix whose rows show the nodes and whose columns show our data. Each entry of matrix RI shows the tendency for the replication of file *n* in node *m*. As the value gets higher, the probability of the replication of file *n* in node *m* will be higher. Because grid system is a dynamic one and is always, changing and we have limited information of the system, every moment we are not able to have accurate information from the whole system and also we can't predict the behaviour of the system. This limited information with the fuzzy system can be used to define some general rules for the system that are always true, so for defining the entries of RI matrix, we use Fuzzy rules.

Through acquiring the matrix entries of each file to each node, a matrix like the following will be formed.

$$RI = \begin{array}{c} \\ n_1 \\ \vdots \\ n_m \end{array} \begin{array}{cccc} f_1 & f_2 & \cdots & f_n \end{array} \\ \left[\begin{array}{cccc} RI_{11} & RI_{12} & \cdots & RI_{1n} \\ \vdots & \vdots & \vdots & \vdots \\ RI_{m1} & RI_{m2} & \cdots & RI_{mn} \end{array}\right]$$

Matrix RI's entries are determined based on Fuzzy methods, are dependent on *Usage-ratio*, *Level*, *Node size*, and *File-size*. *File-size* indicates the *demanded* file size. *Node-size* is the available free space of the node that is considered for accepting replica. Here, it is the aggregation of the free space of the nodes in one cluster.

Since it is possible that the number of demands for files was many in the past but is less now, we also like PHFS[2] use the usage ratio criterion to determine the access history to the file of a node (*Eq.* 3, 4, 5). Consider Table 2.

**Table 2:** The list of Eq. 3, 4, 5 parameters.

| Usage-ratio | Usage ratio |
|---|---|
| Curr-usage | Current iteration usage |
| Prev-usage | Previous iteration usage |

$$usage - ratio = usage - ratio * \frac{Curr - usage}{Prev - usage} \quad (3)$$

If the value of previous usage equals 0, the Formula will be:

$$Usage - ratio = Usag - ratio + Curr - usage \quad (4)$$

If current usage equals 0:

$$Usage - ratio = Usage - ratio/2 \quad (5)$$

Since in the nodes of higher levels, memory capacity and computational power is more than nodes of lower levels, to determine RI, we consider the node level as well in a way that the more a node is in lower level, fewer replications it will accept. *Node-size* in each phase shows the amount of free storage space in each node.

In section 3.1.1, the usage rules for Fuzzy system are defined shortly.

### 3.1.1 Fuzzy Inference system

Fuzzy inference system, is a useful tool for modelling human knowledge, especially if the knowledge about the system is limited to linguistic rules. The aim of this project is making a knowledge base system to infer RI values. Our knowledge about the desired system is very limited, to design a fuzzy system to generate RI values, two fuzzy rules are used which described the RI function in a simple way. The inference system should has a direct relation with *Usage-ratio* and *Node-size* and an opposite relation with *File-size* and the *Node-level*. Fuzzy intersection operators, are not useful for systems with low number of rules. Since our function is not highly sensitive to changes of the value of input variables, average operators can be used instead of intersection operators with lower rules. The fuzzy inference system in form of Eq. 6 uses Mamdani fuzzy product engine using singleton fuzzifier and center of average defuzzifier [22]. Table 3 shows the definition of symbols in Eq. 6 in detail.

$$f(x) = \frac{\sum_{l=1}^{M} y^{-l} \left( \Pi_{i=1}^{n} (\mu_{A_i^l}(x_i)) \right)}{\sum_{l=1}^{M} \left( \Pi_{i=1}^{n} (\mu_{A_i^l}(x_i)) \right)} \quad (6)$$

**Table 3:** The list of Eq. 6 parameters.

| $y^l$ | Center of membership functions for output variable. |
|---|---|
| $\mu_{A_i^l}$ | Membership functions of input variables. |
| $M$ | Number of rules in fuzzy system. |
| $\Pi$ | Fuzzy intersection method. |
| $x_i$ | Input variable. |

The only difference between the proposed fuzzy Inference System, which is shown in Eq. 7, in comparison with Eq. 6, is using fuzzy averaging operator for aggregation of membership functions. Consider Table 4 to know more about the variables in Eq. 7.

$$IF Avg_{Fuzzy} \begin{pmatrix} \text{Level is Low, File} - \text{size is Low, Usage is High,} \\ \text{Node} - \text{size is High} \end{pmatrix}$$
$$THEN\ RI\ is\ High$$

$$IF Avg_{Fuzzy} \begin{pmatrix} \text{Level is High, File} - \text{size is High, Usage is Low} \\ , \text{Node} - \text{size is Low} \end{pmatrix} \quad (7)$$
$$THEN\ RI\ is\ Low$$

$$Avg_{Fuzzy}(a,b) = \lambda\ Max(a,b) + \lambda\ Min(a,b)$$
$$\lambda \in [0\ 1]$$

**Table 4:** The list of Eq. 7 variables.

| File-size | Size of file |
|---|---|
| Node-size | Size of node |
| Level | Level of tree |
| Usage | Usage ratio |

Finally, Eq. 8 is the proposed fuzzy system and Table 5 shows the definition of symbols in detail.

$$f(x) = \frac{\sum_{l=1}^{M} y^l \left( Avg_{Fuzzy_{i=1}^{n}} (\mu_{A_i^l}(x_i)) \right)}{\sum_{l=1}^{M} \left( Avg_{Fuzzy_{i=1}^{n}} (\mu_{A_i^l}(x_i)) \right)} \quad (8)$$

**Table 5:** The list of Eq. 8 parameters.

| $y^l$ | center of membership functions for output variable |
|---|---|
| $\mu_{A_i^l}$ | Membership functions of input variables. |
| $M$ | Number of rules in fuzzy system. |
| Avg | Fuzzy averaging method. |
| $x_i$ | Input variable. |

The input variables of our fuzzy inference system are: *Usage-ratio*, *Node-size*, *Level*, *File-size,* and RI as output variable. Two membership functions of triangle type are LOW and HIGH which considered for each input and output variables. Figure 2 indicates the membership functions with upper and lower limits for all input variables.

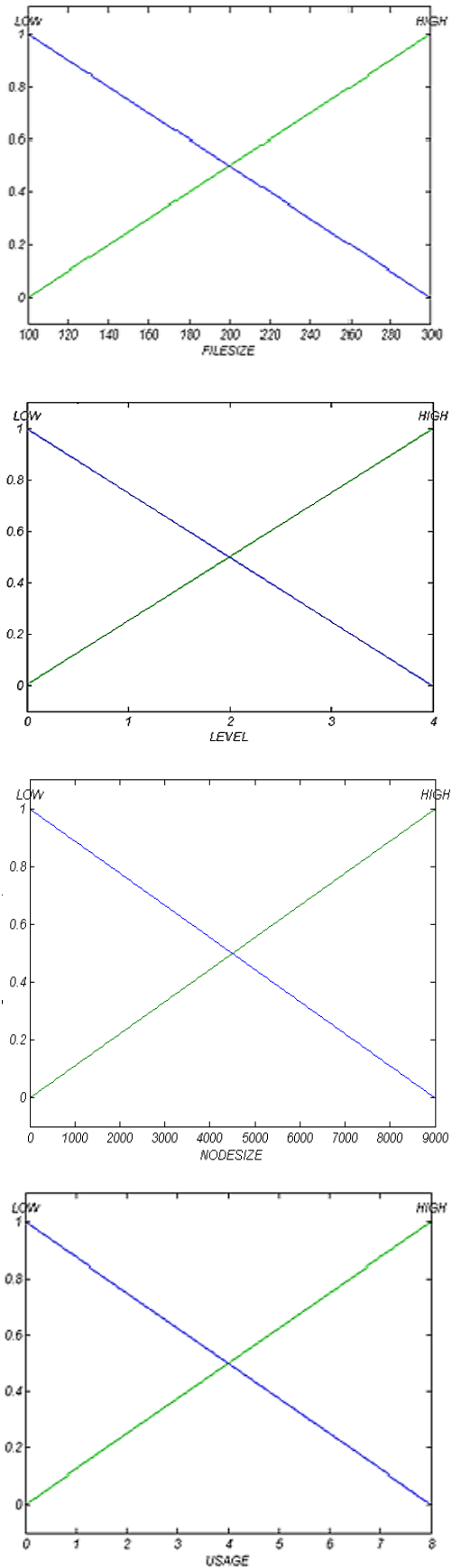

**Figure 2:** Membership functions for all input variables, in the proposed fuzzy inference system.

Similar to input variables, the single output variable has the same membership functions as shown in Figure 3.

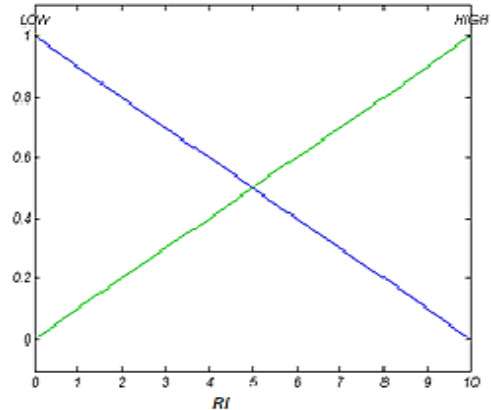

**Figure 3:** Membership functions for output variable RI in proposed fuzzy inference system.

### 3.1.2 Making dependency matrix

In addition to matrix RI, we also need a dependency matrix $n \times n$ in this algorithm, which shows the degree of dependency of one file to other files. Information server is a component in the root of multi-tier tree and gathers all system information. The information server considers files and their information is stored in log files. Information server uses the information in log files and finds the dependency among files by mining the log files using data mining methods. Dependency matrix is made from the gained information in this phase. We suppose that we had these dependencies from the beginning. This paper does not include the ways of using mining techniques.

If we take the dependency ratio as $\beta$, then we have $0 \leq \beta \leq 1$. Dependency of each file with itself equals 1. The entries of main axis of dependency matrix are 1.

### 3.2 PFR algorithm

After each demand, the respective usage history of the demanded file in the node that has demanded is updated and at the end of certain time interval, the RI matrix is recalculated. It should be noted that in each time intervals, several demands can be made. Short time intervals cause an increase in adaptability of the system but impose more overheads to the system. Determination of optimized time interval length for different situation will be studied in future work. At the end of each time interval, the highest entry of RI matrix in which a replication of related file does not exit is chosen. If the chosen entry's usage ratio is higher than a threshold, called $\gamma$, and if there is enough space, algorithm recognizes that file $n$ in node $m$ should be replicated. Here, the amount of $\gamma$ is considered as 2. However, through having a valid data set and learning them via neural network, a dynamic and efficient $\gamma$ is achieved for files with different sizes in different levels. Achieving the optimal value of $\gamma$ is of future works. The respective file is replicated in nodes of different tiers in Multi-Tier Data Grid from root to the node *m* using Fast Spread strategy. In the next phase in dependency matrix, the files whose dependency with file *n* is between 0.5 and 1 are chosen (consider lines 9 to 14 in Figure 7). Suppose that *n* is the replicated file and *n+1* is the file related to it. In the column related to file *n+1*, the existing nodes in the path of respective node for file *n*, i.e. its ancestors, in RI matrix are considered. $S = \{s_1, s_2, ..., s_m\}$ are the nodes on the path of root to the node *m* and $1 \leq j \leq m$. The highest $RI_{s_j \times (n+1)}$ is chosen and we replicate up to that node in the fast spread way (consider lines 15 to 30 in figure 7).

Through this method, not only the files related to *m* are considered but also the history of the file itself, files size, the level of the sites and free available space for replication are taken in to consideration. This is done for all the chosen files of dependency matrix.

The optimizer, is a component in the root of multi-tier tree, and controls the algorithms related to replications. If the file *n* becomes a candidate to be replicated in a node where there is not enough space for the replication of this file, the optimizer provides enough space through deleting the replicas that have lower *RI* in that node until the time we have enough space for replication. PFR, beside the ratio of file dependencies, considers the characteristics of the file itself for replication.

```
PFR Algorithm
1    While(1)
2    {
3      If(time interval finished)
4      {
5        For(each request)
6          "Calculate access rate for the requested file
7          and the requester node."
8      }
9      Define the maximum entry in RI matrix.
10     While( there is not enough replication space)
11     {
12       Remove the replica that has minimum RI.
13     }
14     Replicate the related file of the  maximum entry
       using fast spread strategy.
15     Find the files that their dependency to the chosen
       file is more than 0.5.
16     For(each file that its dependency is more than 0.5)
17     {
18       "Find the maximum RI for each file and the
19       nodes that are in the path between chosen
20       node and the root."
21        While( there is not enough replication space)
22        {
23          Find the replica that has min RI in the node.
24          If (min RI > max RI of dependent file)
25            Go to line 30.
26          else
27            Remove the replica that has minimum RI.
28        }
29     Replicate the file in the node of maximum entry using
30     fast spread strategy.
31       }//end of for
32     }//end of while
```

**Figure 3:** The algorithm of PFR.

### 3.3    Architecture model of PFR

We want to do replication before any demand for files. This algorithm tries to replicate the best file on the most suitable node before demand. As already mentioned, to do this, we need to access history of files and know file dependences on each other.

Our architecture consists of three major parts: Grid Resources, Information Server and Replication Manager (Figure 8).

**Grid Resources:** include computational resources, processors, storage resources and services.

**Information Server:** gathers all system information and has three parts: Clustering phase, File Observation phase and Analysis phase.

**Clustering phase:** as already mentioned, a great number of sites and computational resources, have participated in Grid environment. Therefore, the matrix resulting from all nodes cannot be in practice. We have considered nodes that are close to each other as a cluster. Clustering phase is made up of two parts:
A: Monitoring: in the monitoring phase, the Grid's information is gathered using NWS (Network Weather Service) [27].
B: Clustering: using the gathered information from the observation phase, the algorithm related to clustering is applied.

**File observation phase:** gathers information related to files. This information includes files size and their dependency to each other.

**Analysis phase:** This phase uses the information resulted from file observation phase and applies data mining methods to find the dependency among files, and then stores them in a file called log file.

**Replication Manager:** includes a replicating catalogue in which information related to files is kept. Moreover, replication manager includes a part named Replication optimizer where algorithms related to replications are done. Decision-making about time and method of creation, management and deletion of replicas is a responsibility of this part. The number of demands for each file by the sites of each cluster is kept here. RI matrix and the dependency matrix, which is made of log file in analysis phase, are kept in this place.

As mentioned before, due to great number of sites in grid, instead of discussing sites individually, we discuss clusters and instead of considering all available sites, we will only consider cluster headers related to each cluster. Cluster header is a node in the cluster, which has a higher computational power and storage capacity, compared to other nodes. It has the role of a server and a central unit in the cluster. All of the nodes in the cluster are connected to the cluster header and are in contact with each other via it. Each cluster header separately has a replication manager. RM in each cluster header keeps the information related to all replications present in its own cluster along with the demands for each file by each site.

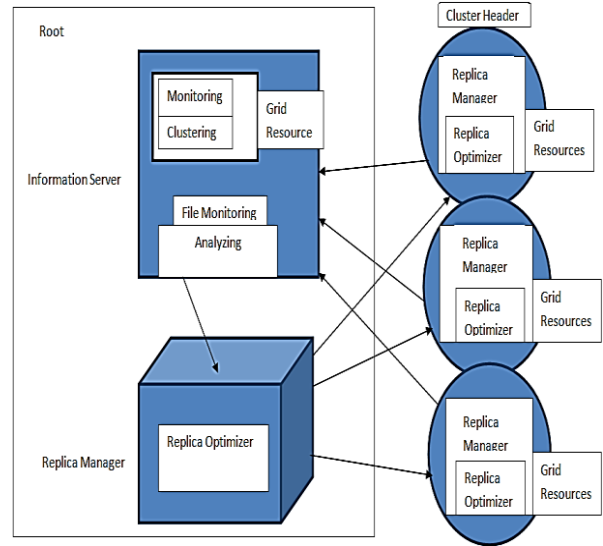

**Figure 4:** System Architecture

Each Replica Manager in each cluster header has a Replica Optimizer. Through use of simple algorithms, RO in each cluster header decides in which site the replication should be placed or from which site it should be deleted and in which site it should be replaced. RO is responsible to control that each cluster has only one replication of each file.

### 3.4 An example of proposed algorithm

In this example we supposed there is enough space for creating replica in each node and there are six cluster headers and five files in the root.

Suppose our RI and dependency matrixes are as follows:

$$RI = \begin{bmatrix} 3.91 & 3.96 & 2.70 & 4.71 & 4.16 \\ 3.20 & 3.40 & 2.78 & 4.10 & 5.00 \\ 3.80 & 3.90 & 2.60 & 4.60 & 4.12 \\ 3.60 & 3.70 & 2.40 & 4.40 & 3.90 \\ 3.91 & 3.96 & 2.70 & 4.70 & 4.16 \\ 4.90 & 3.20 & 5.10 & 3.30 & 4.30 \end{bmatrix}$$

$$\text{dependency} = \begin{bmatrix} 1 & 0.8 & 0.2 & 0.3 & 0.8 \\ 0 & 1 & 0.9 & 0 & 0 \\ 0.7 & 0.7 & 1 & 0.4 & 0.4 \\ 0.5 & 0 & 0.2 & 1 & 0.2 \\ 0.4 & 0.6 & 0.3 & 0.6 & 1 \end{bmatrix}$$

The initial status of the grid is as follows:

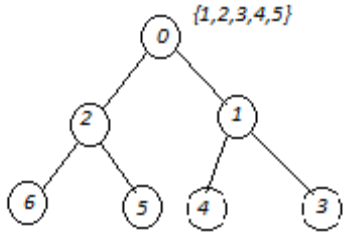

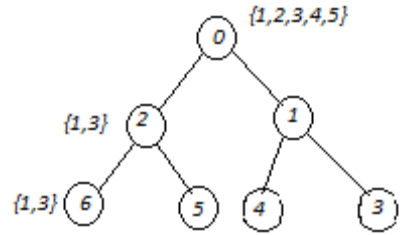

**Figure 7:** After replicating File1, from root to node 6, in fast spread way.

**Figure 5:** Before Replication

After formation of RI matrix, at the end of each time interval, the biggest entry of RI matrix is chosen; it equals $max = RI_{6\times 3}$ here. Since $\gamma$ is 2, if the usage ratio of $RI_{6\times 3}$ is lower than 2, we choose the second higher entry which is $RI_{2\times 5}$ here. The usage ratio of $RI_{6\times 3}$ is 2; then we choose this.

So, file 3 is replicated from root to node 6 in fast spread form. Regarding hierarchical tree, path= {6, 2} (Figure 10).

Also, $RI_{6\times 2} < RI_{2\times 2}$
So file 2 should be replicated from root to node 2 in fast spread way (Figure 12).

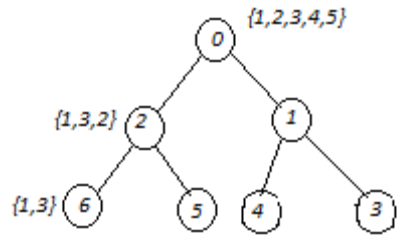

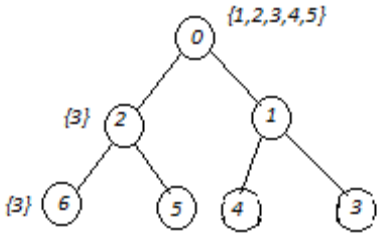

**Figure 8:** After replicating File 2, from root to node 2, in fast spread way.

As you see, by using PFR method, even the dependent files are not replicated only based on their dependency. Their replication is done in the nodes of the tree on the basis of the node and file properties.

**Figure 6:** After replicating File 3, from root to node 6, in fast spread way.

In the next phase in dependency matrix, the files whose dependency with file 3 is between 0.5 and 1 are determined.
Dependency files = {1, 2}
Now, for each of in the Dependency files set, we consider *RI* for each of the nodes that are on the path between root and node 6. For the entry related to each node which is bigger, file is replicated from root to that node in fast spread way.
$$RI_{6\times 1} > RI_{2\times 1}$$
So file 1 should be replicated from root to node 6 in fast spread way (Figure 11).

## 4. Comparison and evaluation

As mentioned before, there are five tiers in the multi-tier architecture we used, with all data being produced at the root. There are 100 nodes in our supposed grid, 14 clusters and 30 files. We do not consider bandwidth between links. The requests for files are generated from all nodes. All three access patterns: Temporal Locality, Geographical locality and Spatial Locality are considered in requests.

In this case, while considering replicating problems, we discuss cluster headers instead of individual sites; this action reduces algorithm complexity from site numbers, to numbers of clusters.

PHFS and Fast spread examine the grid only in the level of leaves and consider only the nodes that demand the files in each level. Requests in these two algorithms are done only from leaves. These two algorithms have better response time compared to PFR. In practice, due to blind creation of the copies in these two algorithms, most of the created replicas are never used and the available resources in grid are wasted as a result of the creation of unused replicas. We can optimize the PFR response time by training the real data set and having more information about the system in order to define better Fuzzy rules. By considering the whole grid and through intelligent creation of the best replicas on the best possible place, PFR uses the resources better than PHFS and CFS. In Figure (13), we can see that the average use of the created replicas at different time intervals in PFR is more than PHFS and CFS. Cascading method is one of the methods that act well in optimized use of resources. We can also see that PFR in the above requests is better than cascading. In this method, we created more optimized copies compared to CFS, PHFS and Cascading and the percentage of the use of the copies increased greatly.

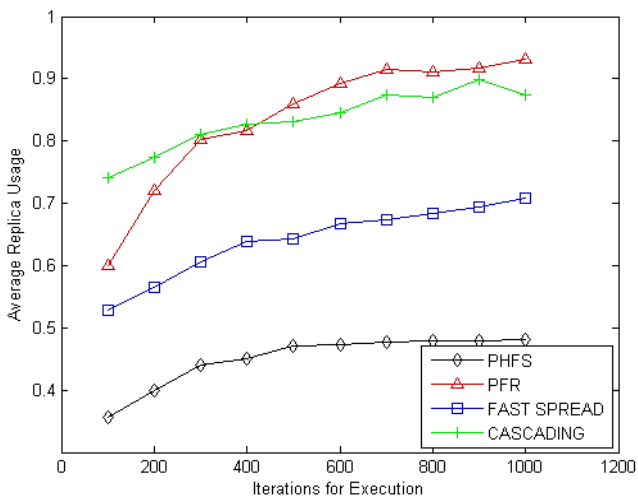

**Figure 9:** comparison Average replica usage for Cascading, PHFS, Fast Spread and PFR

## 5. Conclusions

Considering the dynamic nature of grid, using static thresholds in grid algorithms reduces system efficiency. We can optimize our algorithm by considering the dynamic characteristics of grid and using them in our algorithms. PFR can get better results by having valid data sets and training them via neural networks or by knowing a real system better in order to define better Fuzzy rules.

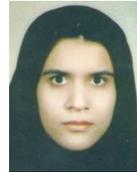

Mahnaz Khojand received the MSA degree from Islamic Azad University of Zanjan. Her research interests include grid computing and cloud.

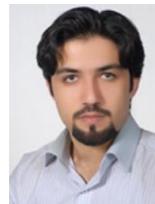

Mehdi Fatan Serj received his B.S. from Islamic Azad University of Shabestar in 2008 and the MSA degree from Islamic Azad University of Qazvin in 2012. He is currently the PhD candidate in URV University in Tarragona. His research interests include grid computing, computer vision, machine learning and robotics.

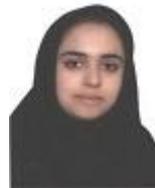

Sevin Ashrafi received the MSA degree from Islamic Azad University of Arak. Her research interests include grid computing.

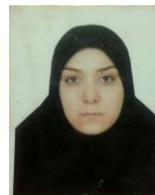

Vahideh Namaki received the MSA degree from Islamic Azad University ofShabestar. Her research interests include grid computing and database.